\documentclass[onecolumn]{elsarticle}

\usepackage{amsmath} 
\usepackage{amssymb}  
\usepackage{amsthm}
\usepackage{graphicx,psfrag}

\newtheorem{definition}{Definition}

\newtheorem{proposition}{Proposition}

\newcommand{\trace}{\mathrm{trace}\,}

\newcommand{\E}{\mathbf{E}\,}

\usepackage{color}

\journal{Physica D}

\begin{document}

\begin{frontmatter}



\title{Finite-time thermodynamics \\ of port-Hamiltonian systems}


\author[JCaddress]{Jean-Charles Delvenne}
\author[HSaddress]{Henrik Sandberg}
\address[JCaddress]{Universit{\'e} catholique de Louvain, Department of Applied Mathematics and Center for Operations Research and
Econometrics (CORE), Louvain-la-Neuve, Belgium. E-mail:  jean-charles.delvenne@uclouvain.be}
\address[HSaddress]{KTH Royal Institute of Technology, ACCESS Linnaeus Centre, School of Electrical Engineering, Automatic Control Lab, Stockholm, Sweden. E-mail: hsan@kth.se}


\begin{abstract}
In this paper, we identify a class of time-varying port-Hamiltonian systems that is suitable for studying problems at the intersection of statistical mechanics and control of physical systems. Those port-Hamiltonian systems are able to modify their internal structure as well as their interconnection with the environment over time.
The framework allows us to prove the First and Second laws of thermodynamics, but also lets us apply results from optimal and stochastic control theory to physical systems.
In particular, we show how to use linear control theory to optimally extract work from a single heat source over a finite time interval in the manner of Maxwell's demon. Furthermore, the optimal controller is a time-varying port-Hamiltonian system, which can be physically implemented as a variable linear capacitor and transformer.
We also use the theory to design a heat engine operating between two heat sources in finite-time Carnot-like cycles of maximum power, and we compare those two heat engines.
\end{abstract}


\end{frontmatter}

\section{Introduction}

Thermodynamics was developed axiomatically in the XIXth century by showing that the empirical observations regarding heat and mechanical work are exchanged between systems could be explained by only four fundamental laws. Making thermodynamics compatible with---or even deducible from---the physical laws of movement of microscopic particles, whether described  by newtonian mechanics or quantum mechanics has been one chief goal of statistical physics, developed by Boltzmann, Maxwell, Gibbs, etc.
The tools of statistical physics allow to model a macroscopical system by a large-dimensional stochastic dynamical system, from which are derived under some assumptions to phenomenological laws for global statistical quantities such as temperature or expected total energy.

Thermodynamics being aimed at studying the exchange of energy between systems and their environment, it is natural that the underlying microscopic system is modelled by an open system, i.e., an equation
\begin{equation}
\dot{x}=f(x,u),
\label{eq:open}
\end{equation}
where $x$ is the state of the system and $u$ is an input provided by the environment, such as an external force or a voltage. When no input is present, the system is said to be autonomous, or isolated: the evolution of the system is not influenced by the environment. One can consider that the environment is itself another open dynamical system, and one may interconnect them in order to get one global autonomous system.

A large part of statistical physics assumes no or infinitesimal exchange of heat with the environment, with probability distributions on the state that are almost constant, avoiding explicit use of open systems theory. However, there are fruitful interactions between open systems theory---also called control theory---and statistical physics, e.g., the dissipation-fluctuation theorem \cite{Kubo}, dissipativity as a control-theory tool \cite{willems72A,willems72B, borkar2003note}, control of stochastic ensembles \cite{Brockett+78,Brockett99}, control thermodynamics \cite{salamon2001principles}, physical interpretation of Kalman filtering \cite{Mitter+05}, port-Hamiltonian systems for irreversible thermodynamics \cite{eberard2007extension}, compartmental systems theory for thermodynamics \cite{Haddad+05}, interconnected thermodynamic control systems \cite{gromov2012stability}, thermodynamic-based Lyapunov functions for distributed dynamical systems \cite{warden1964analysis, favache2009thermodynamics}, model reduction of Hamiltonian systems \cite{barahona+02}, observer's effect in classical statistical physics \cite{Sandberg+11,Sandberg+11b}, etc.

Physical autonomous systems are conveniently expressed in Hamiltonian form $\dot{x}=J(x)\frac{\partial H}{\partial x}$, where $J(x)$ is a skew-symmetric matrix and $H(x)$ is the Hamiltonian, often equal to the total energy of the system. One way to account for the influence of the environment is to assume that the Hamiltonian $H(x,v)$ is dependent on parameters $v$ determined by the environment, e.g., by the presence of an `interaction Hamiltonian' that summarizes the influence of the environment. Another way, essentially equivalent to the first, is the addition to the right-hand-side of a term $g(x)u$ that describes a force field modulated by the environment through the force intensity $u$.

In this paper, we consider both ways alongside, leading to a class of equations
\begin{equation}
\label{eq:ph1}
\dot{x}=J(x,v)\frac{\partial H(x,v)}{\partial x} + g(x,v)u,
\end{equation}
which are called (time-varying) port-Hamiltonian systems. Port-Hamiltonian systems have been introduced in the 1990s for solving deterministic control problems of physical systems, see, for example \cite{vanderschaft-survey2006, Maschke1992, Cervera+07} and the book \cite{Duindam+09}.

The first message of this paper states that port-Hamiltonian systems \eqref{eq:ph1} are a right class of systems for a system-theoretic study of statistical physics, large enough to easily model interesting physical systems and small enough to obey the laws of thermodynamics and be analyzed fruitfully by the techniques of control theory.

The second message illustrates the first and states that the classical problem of the conversion of heat to work can be succesfully formulated and solved as a control problem of a stochastic port-Hamiltonian system in contact with one or several heat baths.  We first find the optimal work extraction from a single-temperature heat bath in a finite time and show to embody the optimal controller, which we call the linear heat engine, in an electric circuit with resistances, capacitance and transformer. We find that the corresponding optimal controller can be interpreted as implementing a zero-temperature low resistance, or in other words a Maxwellian demon \cite{maxwell1897theory, leff2010maxwell} that extracts as much work as desired from a single source of heat. Explicit expressions for the finite-time work and heat are derived. As this optimal linear heat engine is not cyclic, we look for Carnot-like cyclic solutions operating between several temperatures and see the impact of the dynamical time constants of the system on efficiency and power of the cycle. This offers a dynamic interpretation of some classical results of endoreversible and finite-time thermodynamics \cite{andresen1983finite, orlov1990power, salamon2001principles,hoffmann2008introduction}.

Among the closest work in the literature are \cite{Brockett+78}, from which we borrow the illuminating example of time-varying capacitance, \cite{Brockett99}, in which bilinear systems are studied from a stochastic control point of view with results for finite-time Carnot cycles illustrated on a piston-and-gas nonlinear example, \cite{delvenne+07}, which formalizes a dissipativity and work extraction theory for the time-varying linear systems, and \cite{CDC07HeatEngines}, which analyses some examples of linear heat engines. Other recent proposals for a physical implementation of Maxwell's demon, although of a very different nature than ours, are \cite{mandal2012work, strasberg2013thermodynamics}.

The paper is organized as follows. In Sections \ref{sec:port-ham}--\ref{sec:diss-port-ham} the class of port-Hamiltonian systems is discussed, with a special focus on linear (Section \ref{sec:lin-p-h}) and scalar linear (Section \ref{sec:time_varying_cap}) systems. How they obey the First and Second Laws of thermodynamics, thus Carnot's theorem, is explained in Section \ref{sec:thermo}. A study of non-cyclic and cyclic optimal linear heat engines is proposed in Section \ref{sec:finite-time}.

\section{Lossless port-Hamiltonian systems}
\label{sec:port-ham}
This paper seeks to analyze work extraction of thermodynamic systems as a stochastic optimal control problem on  fundamental physical systems. We must therefore choose a class of open systems that is large enough to allow the convenient modelling of any open system from first principles of classical physics, and small enough so that  bounds can be derived on the achievable performance of the relevant control tasks on such systems, e.g., Carnot's theorem.

We propose the so-called \emph{port-Hamiltonian systems} as such a suitable class. Port-Hamiltonian systems have been developed since the 90s by the open systems community as a mean to formulate control problems on mechanical and electrical systems. In particular, an intrinsic formulation has been initiated by \cite{Schöber+12}, although in this paper we shall express the dynamics in coordinate form. The following formal definition is followed by some explanatory comments and reminders.

\begin{definition}
In this paper, we call a system \emph{lossless port-Hamiltonian} if it obeys equations of the form
\begin{equation}
\dot{x}=J(x,v) \frac{\partial H(x,v)}{\partial x} + g(x,v) u
\label{eq:ph}
\end{equation}
where 

\begin{itemize}
	\item the \emph{symplectic form} $J(x,v)$ is skew-symmetric (i.e., $J+J^T=0$), invertible and closed in $x$ (i.e. satisfying
 $\partial_i K_{jk}+\partial_k K_{ij}+\partial_j K_{ki}=0$ for $K=J^{-1}$);
	\item $g(x,v)$ \emph{derives locally from a gradient} in $x$, i.e. can be expressed as $g(x,v)=J(x,v)\frac{\partial G(x,v)}{\partial x}$ for some scalar `potential' $G(x,v)$; 
\item $u(t)$ (called the \emph{linear input}) and $v(t)$ (called the \emph{nonlinear input}) are time-varying parameters representing the influence of the environment on the dynamics of the system
\end{itemize}
 The \emph{linear output} $y$ is defined as $y=g^T(x,v) \frac{\partial H(x,v)}{\partial x}$.  The \emph{nonlinear output} $z$ is defined as $\frac{\partial H(x,v)}{\partial v}$.
\label{def:1}\end{definition}

The matrix $J$ being skew-symmetric, invertible and closed ensures that in the absence of any external interference, i.e. when $u=0$ and $v$ is constant, the system is a mere Hamiltonian system. By Darboux's theorem, those are exactly the systems that can be expressed by the familiar Hamilton's equations $\dot{q}=\partial H / \partial p, \quad \dot{p}=- \partial H / \partial q$ by a local change of variables $(q(x),p(x))$. We can then interpret the coordinates $q_i$ as (generalised) positions, associated to (generalised) momenta $p_i$. In particular, the state space is always even-dimensional.

An input $u_i$ may for instance be a force, and the corresponding output $y_i$ the speed of the point on which the input force is exerted, or the other way around. The input $u_i$ may as well be a current through a two-terminal port, and the output $y_i$ a voltage across the port, or the other way around. If the nonlinear input $v$ is a constant, then it is easy to see that we can express the energy balance as $\dot{H}=\sum_i u_i y_i$, so that $u_iy_i$
 may be interpreted as the power flowing through the port $(u_i,y_i)$ (see \cite{willems2010terminalsIEEEMag} for more careful conditions to which this interpretation is physically unambiguous). 
 If $v$ is time-varying, $\dot{H}=\sum_i u_i y_i + \sum_j \dot{v}_j z_j$, thus justifying the form of the nonlinear output $z_j$ as $\frac{\partial H}{\partial v_j}$, corresponding to the nonlinear input $v_j$.

We highlight, to avoid any confusion, that the definition above of lossless port-Hamiltonian systems slightly differs from the most common definition found in the control literature, see, for example \cite{vanderschaft-survey2006, Maschke1992, Cervera+07, duindam2009modeling}, where in particular only linear inputs are considered, $J$ is not required to be invertible and closed, and any force field $g(x)u$ is allowed, making it possible to model phenomenological non conservative forces.

We may interconnect two port-Hamiltonian systems $(x_i, u_i, y_i)$ ($i=1,2$) on the port $(u_i, y_i)$ in feedback, making one system's input the other system's output. More precisely, we may have for instance the interconnection equations $u_1+y_2=0$, if $u_1$ and $y_2$ represent equal and opposite forces,  and $u_2=y_1$ represent a same speed. In an electrical context, the currents check $u_1+y_2=0$ by Kirchhoff's Law and the voltage between two same points are denoted $y_1=u_2$. In all cases, such a lossless interconnection ensures that $\dot{H}_1(x_1)=-\dot{H}_2(x)=y_1^Tu_1$, while the total energy remains constant. It is easy to see that the total dynamics is described by a Hamiltonian system with energy $H_1+H_2$ and the symplectic form $J=\begin{pmatrix} J_1 & -g_1g_2^T \\ g_2g_1^T & J_2 \end{pmatrix}$. This symplectic form is closed indeed, due to the fact that $g_i$ both derive from a gradient.

Linear and nonlinear controls, representing apparently two different ways
to act on the system, are essentially equivalent, as we now argue.

We first observe that a linear input can be seen as a particular nonlinear input. Since the force $g(x,v)u$ is a gradient $J \frac{\partial G}{ \partial x}$, Equation (\ref{eq:ph}) can be rewritten as a Hamiltonian system of Hamiltonian $H(x,v)+u G(x,v)$. Although it is obviously equivalent in terms of evolution of the state space, it must be underlined that it leads to a different balance of energy. Indeed, $u$ is now seen as a `nonlinear' input with corresponding output $z=\partial(H+uG)/\partial u= G$, and the Hamiltonian $H+uG$ varies at the rate $\dot{u} G$. By contrast, the linear input $u$ and the corresponding  linear output $y=g^T \partial H/\partial x$ in Equation (\ref{eq:ph}) leads to a rate $u \dot{G}$ for the Hamiltonian $H$. A concrete example of this difference would be a mass $m$ of height $x$ in a gravitational field of strength $u$. One may either consider the mass as a port-Hamiltonian system of Hamiltonian $m \dot{x}^2/2$ subject to an external force $u$, or a Hamiltonian system of Hamiltonian $m \dot{x}^2/2 + m u x$, which amounts to including the gravitational field inside the system. Although these two views are equivalent in terms of equation of motions, they interpret the `internal energy' of the system and its variation differently. The choice is a matter of convenience. Of course if at some initial and final times $0$ and $T$ the system is isolated from its environment, with $u(0)=u(T)=0$, the two interpretations coincide, the Hamiltonian is defined unambiguously in these moments, and it is important that the variation $H(T)-H(0)$ is predicted equally by the two models. It is indeed the case as shown by an integration by part. Similarly, any cyclic boundary conditions ensuring that $u(0)G(0)=u(T)G(T)$ will give a same energy
balance over a cycle. Therefore any optimal energy extraction problem with these kinds of boundary conditions, as considered later in this paper, are well-posed. Restrictive conditions under which the instantaneous power transfered to a system is defined at all times unambiguously are given in \cite{willems2010terminalsIEEEMag}; those conditions are satisfied in the case of usual electric circuits such as those we use as examples in this paper.

Conversely, one may for every nonlinear input $v$ add two states $v$ and $m\dot{v}$, and create a corresponding linear input $u$ with the state space equations
\begin{equation}
 \begin{pmatrix}
      \dot{v} \\
      m \ddot{v} \\
    \end{pmatrix}
    =
     \begin{pmatrix}
                    0 & 1 \\
                    -1  & 0 \\
    \end{pmatrix}
    \begin{pmatrix}
                       \partial H/ \partial v \\
                      \dot{v} \\
    \end{pmatrix}
       +
    \begin{pmatrix}
                     0 \\
                     u \\
    \end{pmatrix},
    \label{eq:eps}
\end{equation}
and an augmented Hamiltonian $H'=H+m \dot{v}^2/2$, for any constant $m$. We suppose, for consistency, that the rest of the system is also in canonical coordinates, or with constant $J$. The corresponding output is $y=\dot{v}$. Thus for every trajectory
$v(t)$ we can find a corresponding linear input trajectory $u(t)=\partial H/ \partial v + m \ddot{v}$ with the same effect on the state variables $x$. As $m$ vanishes, the work $yu$ exerted on the system is arbitrarily close to $\dot{v}\partial H/ \partial v$.
For instance, a time-varying capacitance $C$ in an electric circuit contributes to the Hamiltonian with a term $\frac{q^2}{2C}$ associated with the electric charge $q$. The capacitance $C$ is a nonlinear input that can extract energy from the system through its time variation. On the other side we may consider the detailed mechanism through which the capacitance is varied. For example, a capacitance may consist of two plates of mass $m$, the distance between which can be varied. In this case, the nonlinear input is the distance $v$, which is practically modified through the application of a force $u$, following Equation \ref{eq:eps}. But we may also create a time-varying capacitance by other means, such as changing the dielectric between the plates. Therefore, not only is the nonlinear input a simpler choice as it leads to more compact equations, but it ensures that any bound on the control performance obtainable from that system will be independent from the particular physical implementation of the nonlinear control.

In summary, the class of lossless port-Hamiltonian systems contains the Hamiltonian systems, is invariant under interconnection and allows two channels of interaction with the environment, through linear and nonlinear inputs/outputs. While those channels are largely redundant in principle, they offer a flexibility to model various situations easily. It is a reasonable stance to believe that
the fundamental control of an environment over a system is through a linear control $u$, e.g. a force field, while the general nonlinear control $v$ influencing for instance the Hamiltonian is only a phenomenological expression allowing to model an interaction at a higher level than by the detailed description of the interaction mediating the influence of the environment.

\section{Dissipative port-Hamiltonian systems}
\label{sec:diss-port-ham}

Sources of energy dissipation, such as friction or electric resistance, are indispensable components for a convenient modelling of macroscopic situations. The most common model of dissipation involves a linear direct relationship between input and output (between force and speed, or between current and voltage): $u=ry$. It is known that this simple relationship can be implemented arbitrarily well by a many-dimensional fundamental linear system \cite{nyquist28,caldeira+83,Sandberg+11}. The initial state of the many-dimensional
 internal state can only be described by a probability distribution. The effect of this random initial distribution over a great many degrees of freedom translates into an additive noise. There are fairly good theoretical and empirical reasons to shape the effect of this noise as
 \begin{equation}
u=ry+ \sqrt{2rT}n(t),
\label{eq:johnson}
\end{equation}
where $n(t)$ 
is a Gaussian white noise of unit intensity (the `derivative' of a Brownian motion) and $T$ is a constant called temperature, that scales the amplitude of the noise. The noise is often called Johnson-Nyquist noise.

In this paper, we limit ourselves to linear resistors because a general noise model for nonlinear
resistors is not known and seems out of reach
(see \cite{Gupta78} and references within for partial results and a discussion). We do however allow a
resistance $r(x,v)$ that depends on the state or external environment.

Assume a port-Hamiltonian system with a resistance $r$ as in Equation \eqref{eq:johnson} connected between a scalar input $u$ and a scalar output $y=g^T \frac{\partial H}{\partial x}$. Then the global equation reads

\begin{equation}
\dot{x}= (J - r g g^T) \frac{\partial H}{\partial x} + \sqrt{2rT}g n
\label{eq:dissip1}
\end{equation}

As we may add several resistances to several ports, one arrives at the following general form, which could also be called an `open Langevin equation'.

\begin{definition}
A dissipative port-Hamiltonian system is of the form
\begin{equation}
\dot{x}= \left(J(x,v) - \sum_i R_i(x,v)\right) \frac{\partial H(x,v)}{\partial x} + g(x,v) u + \sum_i \sqrt{2R_i(x,v)T_i}n_i(t)
\label{eq:dissip2}
\end{equation}
where $v$ is a nonlinear input vector, $u$ is a linear input vector, $J$ and $g$ satisfy the conditions of a lossless port-Hamiltonian system, $R_i(x,v)$ is a symmetric nonnegative definite matrix, a square root of which is denoted $\sqrt{R_i}$,  
$T_i$ is the temperature of $R_i$, and $n_i$ are independent unit intensity Gaussian white noises. The outputs are as in Definition \ref{def:1}.
\end{definition}

It should be said here whether we understand the above stochastic differential equation in the It\={o}'s or Stratonovich's sense. While It\={o}'s calculus is popular in the control community because of its causality properties, Stratonovich calculus is often more adapted for physical situations. The two intepretations coincide whenever $R_i(v)$ is independent on $x$.

\section{Linear port-Hamiltonian systems}
\label{sec:lin-p-h}
It is customary to linearise an autonomous dynamical system in the vicinity of a stable fixed point in order to understand its behaviour. By (marginal) stability, the linear approximation remains approximately valid at all times in a certain neighbourhood of the origin.

It is common for an open system to have stable fixed point $x=0$, which corresponds to a local minimum of the Hamiltonian, when the system is isolated from the environment (i.e., whenever $u=0$ and $v$ is constant). We may linearize all trajectories around this equilibrium, replacing $J(x,v)$ by $J(0,v)$, $R(x,v)$ by $R(0,v)$, $g(x,v)$ by $g(0,v)$ and  most importantly $H(x,v)$ by $H(0,v) + \frac{\partial H(0,v)}{\partial x} x + \frac{1}{2} x^T  \frac{\partial^2 H(0,v)}{\partial x^2} x$.
We can assume $H(0,v)=0$ without loss of generality. As $x=0$ is a local minimum of $H(x,v)$ for all $v$,  $\frac{\partial H(0,v)}{\partial x}$ is zero. Therefore, one can consider a Hamiltonian of the form $\frac{1}{2} x^T \Sigma(v) x$, for a positive definite $\Sigma(v)$. As we shall see, this particular family of linear systems covers a large number of practically interesting cases.

The global dynamics is therefore of the form
\begin{equation}
\dot{x}= (J(v) -R(v))\Sigma(v) x + g(v) u + \sqrt{2R(v)T}n(t),
\label{eq:lintv1}
\end{equation}
with linear output $y=g^T\Sigma x$.

It has not been stressed so far that one system has a whole family of representations by differential equations, related to one another by change of coordinates. It is sometimes convenient to consider time-varying change of coordinates $\tilde{x}= P(t) x$,
leading to a new equation in the $\tilde{x}$ coordinates:

\begin{equation}
\frac{d}{dt}\tilde{x}= (\tilde{J}(v) + \tilde{M}(v) -\tilde{R}(v))\tilde \Sigma(v) \tilde{x} + \tilde{g}(v) u + \sqrt{2\tilde R(v)T}n(t),
\label{eq:lintv2}
\end{equation}
where $\tilde{J}$ and $\tilde{M}$ are defined as the skew-symmetric and symmetric parts of $P J P^T+ \dot{P} \Sigma^{-1} P^T$, while $\tilde{g}=Pg$ and $\tilde{R}=PRP^T$.  The linear output is $y=\tilde{g}^T \tilde{\Sigma}\tilde{x}$.

The term $\tilde{M}(v)$ plays the same role as the dissipation term $-\tilde{R}(v)$, except that it is not necessarily negative definite and it is not matched by a random fluctuation term. In other terms it acts as a positive or negative zero-temperature resistance, that represents a loss or gain of energy by the system. As we shall see later on, this is associated to work (mechanical or other) performed on the environment. Energy-normalizing coordinates, which makes the energy form $\tilde \Sigma(v)$ equal the identity at all times by choosing $P^T(t) P(t)=\Sigma(v(t))$, are particularly convenient if they exist, as seen in the later sections.  In case $\tilde \Sigma(v)$ is only nonnegative definite then we normalize it to a diagonal zero-one matrix $D$ by choosing 
$P^T(t) D P(t)=\Sigma(v(t))$. In this paper we usually consider the positive definite case.

\section{Scalar linear systems}
\label{sec:time_varying_cap}
\begin{figure}[tb]
\centering
  \includegraphics[width=0.95\hsize]{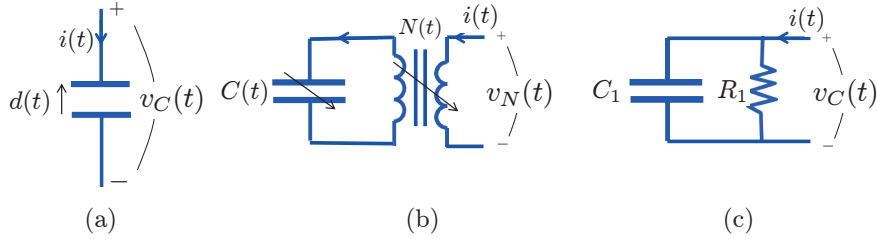}\\
  \caption{(a) A time-varying capacitor, with variable distance $d(t)$ between the plates. (b) The time-varying capacitor interconnected to a time-varying transformer. (c) An RC-circuit with temperature zero. With proper choice
  of $N(t)$ and $C(t)$, the circuits (b) and (c) can be made externally equivalent.}\label{fig:varying_cap}
\end{figure}

Most of the examples will be drawn from the linear systems detailed above, hence their importance.
Let us consider in more detail the time-varying capacitor in Figure~\ref{fig:varying_cap}-(a), whose capacitance $C$ can be modified at will by, e.g., moving the plates of the capacitance and acts as a nonlinear input $v_1=C$.
 The linear input is the current,  $u=i$, and the linear output is the voltage, $y=v_C$. We can choose the state to be  the charge $q$,  the voltage $v_C$, or $x = q/\sqrt{C}=q/\sqrt{v_1}$, with corresponding equations
\begin{equation} 
\begin{aligned}
\dot{q} &=0q + u, & y&=q/v_1 & \text{with}\quad &  H= \frac{1}{2v_1} q^2; \\
\dot{v}_C &= -\frac{\dot{v}_1}{v_1}v_C + \frac{1}{v_1} u, & y&=v_C  & \text{with}\quad  & H= \frac{1}{2}v_1 v_C^2; \\
\dot x &=
-\frac{\dot{v}_1}{2v_1}  x + \frac{1}{\sqrt{v_1}} u,
& y&=\frac{1}{\sqrt{v_1}}  x & \text{with}\quad &  H= \frac{1}{2}  x^2.\\
\end{aligned} \label{eq:capacitor}
\end{equation}

Note that these equations involve only one state, with a non-invertible $J$, since here $J=0$. One may artificially add a dummy state variable, e.g., $\int q $, which has no influence in the Hamiltonian or in the input-output relationship. This would allow a proper Hamiltonian structure with an even-dimensional state space.  For simplicity we keep the above one-dimensional systems. We will in the following choose to work with the third energy-normalizing state-space representation satisfying the energy balance
\begin{equation}
\dot H = -M(v_1) x^2 + yu \quad\text{with} \quad  M(v_1) = \frac{\dot v_1}{2v_1}.
\label{eq:dotH_var_cap}
\end{equation}
The mechanical work extraction rate from the moving plates of the capacitor is $M(v_1)x^2$, while
the product $yu$ is the electrical power into the capacitor.

We find it useful to connect the above time-varying capacitor to a time-varying ideal lossless transformer \cite{Anderson+65} with
varying turns ratio $N > 0$. The turns ratio is our second nonlinear input, $v_2=N$. The system is illustrated in Figure~\ref{fig:varying_cap}-(b) and the model becomes
\begin{equation}
\begin{aligned}
    \dot{x} & = -\frac{\dot v_1}{2v_1}x + \frac{v_2}{\sqrt v_1}u, \quad
    y = \frac{v_2}{\sqrt v_1}  x,\quad \text{with}\quad H = \frac{1}{2} x^2,
\end{aligned}
\label{eq:captransdev}
\end{equation}
where $u=i$, $y=v_N$, $v_1>0$, and $v_2> 0$.
This time-varying circuit can be used to implement a large class of first-order linear time-varying systems, as
stated in the following proposition, proved in the Appendix.
\begin{proposition}
\label{prop:approx}
The input-output map of the first-order linear time-varying system
\begin{equation}
\begin{aligned}
\dot p & = a(t)p + b(t)u, \quad
y = c(t)p,\quad p(0)=0,
\end{aligned}
\label{eq:genlin}
\end{equation}
where $a\in\mathcal{C}^0$, $b,c\in\mathcal{C}^1$, can be exactly implemented using
the port-Hamiltonian system \eqref{eq:captransdev} with $x(0)=0$ if and only if $b(t)c(t)> 0$ for all $t$. An implementation is obtained with the
nonlinear inputs
\begin{align*}
    v_1(t)  & =e^{-2\int_0^t a(s) ds} \frac{b(t)c(0)}{c(t)b(0)}v_1(0), \quad
    v_2(t) =  \sqrt{b(t)c(t)v_1(t)},
\end{align*}
for arbitrary $v_1(0)>0$.
\end{proposition}

To show how the circuit can be used, let us use it to implement an ideal \emph{time-invariant} RC-circuit, see Figure~\ref{fig:varying_cap}-(c), where the resistor has temperature zero and exhibits no Johnson-Nyquist noise.
Assuming the resistance is $R_1>0$, $u=i$, and $y=v_C$, the model (in energy-normalizing coordinates) becomes
\begin{equation}
\dot p = -\frac{1}{R_1C_1} p + \frac{1}{\sqrt{C_1}}u, \quad y = \frac{1}{\sqrt{C_1}} p.
\label{eq:RC-circuit}
\end{equation}
Thus $a=-1/(R_1C_1)$ and $b=c=1/\sqrt{C_1}$, and let us denote the time-constant of the circuit by $\tau_1 = R_1 C_1$. Then we should according to Proposition~\ref{prop:approx} choose the following nonlinear inputs for the  port-Hamiltonian implementation \eqref{eq:captransdev}
\begin{equation*}
v_1(t)= e^{2t/\tau_1}v_1(0), \quad v_2(t) = \sqrt{v_1(t)/C_1}.
\end{equation*}
Physically this corresponds to a compression of the capacitor plates and that mechanical energy is extracted in a way to resemble the
dissipation of the RC-circuit. To further illustrate the flexibility of the time-varying circuit, note that we can easily also implement an active filter with a \emph{negative} resistor $-R_1<0$ by just changing the nonlinear inputs
to
\begin{equation*}
v_1(t)= e^{-2t/\tau_1}v_1(0), \quad v_2(t) = \sqrt{v_1(t)/C_1}.
\end{equation*}
Physically this corresponds to pulling the capacitor plates apart which requires the mechanical work injection rate $|w| = x^2/\tau_1$.

Remark that not only the input-output map of the RC-circuit is replicated by the time-varying circuit, but also the amount
of energy stored in the capacitor, since $x(t)=p(t)$ for all $t$. Hence, in this sense the time-varying port-Hamiltonian system is both externally and internally equivalent to the RC-circuit.

That  work extraction or injection can be interpreted as an equivalent $RC$ circuit will be useful in later sections to analyze heat engines, and is further formalized for general systems in the following section.

\section{The First and Second Law of thermodynamics}
\label{sec:thermo}

The class of port-Hamiltonian systems obeys the First and Second Law of thermodynamics, as we now detail.

Consider a port-Hamiltonian system with a random state $x$. The \emph{internal energy} is defined as
the expected Hamiltonian $U=\E_x H(x,v)$. For a lossless system steered by inputs $u,\, v$, the variation of internal energy $\dot{U}$ is interpreted as the \emph{power} or \emph{work rate} $w$ performed by the system on the environment:
$$
w=-\E_x u^T y - \E_x \dot{v}^T  \frac{\partial{H}}{\partial{v}}.
$$
For a dissipative system as described by Equation \ref{eq:dissip2}, the energy balance is written, following the rules of (It\={o}'s) stochastic calculus:
$$
\dot{U}= -w - \E_x \frac{\partial H}{\partial x}^T R \frac{\partial H}{\partial x} + T \E_x \trace \, R \, \frac{\partial^2 H}{\partial x^2}.
$$
Note that using Stratonovich's calculus would add a term $T \sum_{ijk} \partial H/ \partial x_i \partial S_{ij}/\partial x_k S_{kj}$ with $S$ being a square root of $R=SS^T$. This would modify the expression for heat $q$ in the subsequent developments of this section. This term vanishes if $R$ does not depend on $x$, as is the case in all our examples. For simplicity we keep the It\={o} form below. 

The term  $- \E_x \frac{\partial H}{\partial x}^T R \, \frac{\partial H}{\partial x}$ represents the dissipation through the resistive parts, while
$T\E_x \trace R \, \frac{\partial^2 H}{\partial x^2}$ is due to the fluctuation. Together they form the \emph{heat} rate:
$$
q=- \E_x \frac{\partial H}{\partial x}^T R \frac{\partial H}{\partial x} + T \E_x \trace \, R \, \frac{\partial^2 H}{\partial x^2},
$$
so that the energy balance, or \emph{First Law} of thermodynamics, is written simply as
$$
\dot{U}=q-w.
$$
A linear port-Hamiltonian system under the energy-normalizing coordinates $x$, obeying
$$
\dot{x}=(J(v)+M(v)-R(v))x+g(v) u + \sqrt{2R(v)T}w,
$$
has internal energy $U=\frac{1}{2} \E_x x^Tx = \frac{1}{2} \trace \, X $ for which the First Law becomes
\begin{align*}
\dot{U}&=q-w,\quad
w= - \E_x u^Ty - \trace \, M(v)X  , \quad
q= - \trace \, R(v)X+ T\, \trace \, R(v) ,
\end{align*}
where $X:= \E_x xx^T$ is the second moment of $x$.

As we observed in the previous section for the scalar case, we see that the extraction of work is formally undistinguishable from positive or negative dissipation elements,
thus can be represented e.g. by time-varying zero-temperature resistances in  an electric circuit. This key intuition will be used in the next section for the convenient design and analysis of heat engines.

Kelvin's statement of the \emph{Second Law} states that one cannot extract  work from a unique source of heat through a cyclic process. This can be proved rigorously for port-Hamiltonian systems with classical arguments, which we overview briefly. A way to prove it is to introduce a quantity called entropy, denoted $S$, which is the differential Shannon entropy of the probability distribution of the state relative to the measure $\mu$ defined by the symplectic structure associated with the $J$ matrix. The quantity is finite only when the state probability is characterized by a probability distribution $\rho(x,t)$ with respect to $\mu$, in which case it is equal to $S = - \E_x(\ln \rho)$. Using the fact that the Kullback-Leibler divergence of two initial probability measures for the same Markovian process is non-increasing \cite{coverthomas} and the fact that the Gibbs distribution $\rho(x) \propto \exp(-H(x,v)/T)$ is a stationary probability measure for one heat bath $T$, it is classic to derive the celebrated \emph{Clausius inequality}
$$
\dot{S} \geq \sum_i \frac{q_i}{T_i},
$$
where $q_i$ is the heat rate exchange with the heat bath of temperature $T_i$.
From there it is quite elementary to derive Carnot's theorem, which states that the efficiency $\eta= \oint w/\oint q_{\textrm{hot}}$ of a system having access to heat baths $T_{\textrm{hot}} > T_{\textrm{cold}}$ is bounded by $1-T_{\textrm{cold}}/T_{\textrm{hot}}$. Moreover this efficiency can be attained arbitrarily close by cycles that access at most one bath at a time and for which $\sup_t |q(t)|$ is arbitrarily small, i.e, the cycles are infinitely slow.

Matrix-theoretic proofs of this for the linear port-Hamiltonian systems can be found in \cite{delvenne+07}, generalizing \cite{Brockett+78}.
In \cite{Brockett99}, a proof for bilinear systems is provided along with interesting efficiency bounds on finite-time cycles with the tools from stochastic control theory. Our aim now is to proceed further into this direction, although our starting point is different in that we use the port-Hamiltonian framework and optimal linear-quadratic control theory.

\section{Finite-time transformations}
\label{sec:finite-time}

In this section, we first describe a simple non-cyclic optimal linear
heat engine that extracts work from a single heat source. Then we will discuss its finite-time implementation using physical components, and its relation to Carnot heat engines, which are known to achieve the optimal thermodynamic efficiency.

\subsection{An Optimal Linear Heat Engine}
\label{sec:lqr}

\begin{figure}[tb]
\centering
\psfrag{v}[][]{$y$}
\psfrag{j}[][]{$u=i$}
  \includegraphics[width=0.6\hsize]{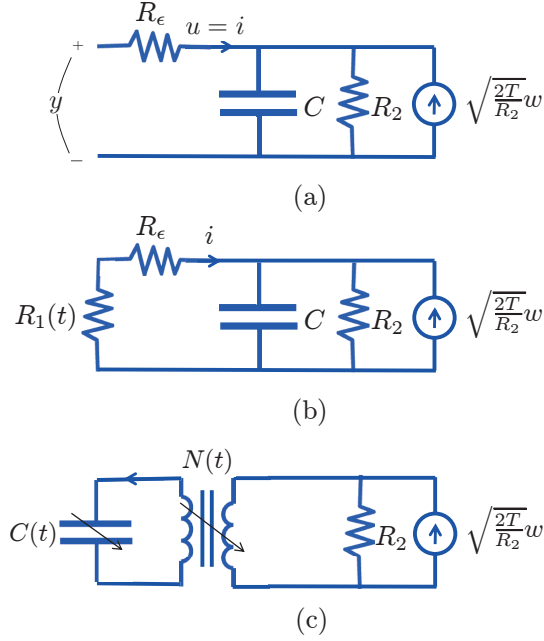}\\
  \caption{Circuits for the optimal linear heat engine. (a) The open single-temperature system from which work is to be extracted. (b) The linear heat engine (or Maxwell's demon) $R_1(t)$, is added to control the system in order to extract work. (c) A time-varying lossless implementation of the system in (b).}\label{fig:LQR_circuit}
\end{figure}

Let us consider a resistor $R_2$ of temperature $T$, whose effect can be modelled by a parallel source of random white noise current $\sqrt{2T/R_2} n$, see \eqref{eq:johnson}. Looking at the frequency domain, it is well known that every frequency band $\Delta f$ carries a power $4R_2T\Delta f$. By connecting a zero-temperature resistance $R_1=R_2$ in parallel, one can dissipate a power $T \Delta f$, therefore an infinite power over all frequencies, through the resistance $R_1$.  Remember that a zero-temperature resistance can be implemented by a capacitor with moving plates, see Section~\ref{sec:time_varying_cap}, so that this `dissipated' energy is actually work extracted from the system.  This apparent ability for a hot resistor to exchange an infinite amount of thermal energy with the environment is sometimes called the \emph{ultraviolet catastrophe}. Of course, this diverging power only betrays the limit of the Johnson-Nyquist white noise model. In reality, high frequencies power vanish due to fundamental reasons (quantum cut-off \cite{nyquist28}) or engineering constraints (limited heat conductivity).
We assume therefore that there is a given capacitance $C$ in parallel with $R_2$, which filters out the high frequencies of the noise, with cut-off frequency $1/\tau_2$, where $\tau_2=R_2C$ is the time constant of the $R_2 C$ circuit, see Figure~\ref{fig:LQR_circuit}-(a). Intuition may therefore suggest that we can retrieve
at best a mechanical power of the order of $T/\tau_2$. In this section, we show under what condition it is true, and how  to extract a maximum amount of useful work from the hot $R_2 C$ circuit within a time $t_f$.

Using energy-normalizing coordinates $x=\sqrt{C}v_c$ where $v_c$ is the voltage across the capacitor, we have the model
\begin{equation}
\begin{aligned}
\dot{x} & = -\frac{1}{\tau_2} x  + \frac{1}{\sqrt{C}}u  + \sqrt{\frac{2T}{\tau_2}}n, \quad \E x(0)=0,\,\E x(0)^2 = T, \\
y & = \frac{1}{\sqrt{C}}x + R_\epsilon u,
\end{aligned}
\label{eq:cap_heat_source}
\end{equation}
where $H = \frac{1}{2}Cv_C^2 = \frac{1}{2}  x^2$ is the Hamiltonian. The input is the injected current
$i$ and the output $y$ is the voltage across the circuit, see Figure~\ref{fig:LQR_circuit}-(a). The resistor $R_\epsilon$ of temperature zero represents
inefficiencies in the work extraction mechanism, and we will let it tend to zero later. Alternatively one can interpret $R_\epsilon$ as
losses in the interconnecting wires. We assume that $R_\epsilon$ has zero temperature, because a Johnson-Nyquist noise
in $R_\epsilon$ would again provoke an ultraviolet catastrophe and an infinite power, thus deteriorating the accuracy of the
model rather than improving it.

\begin{figure}[tb]
\centering
    \psfrag{R1(t)}[][][0.9]{$R_1(t)$}
    \psfrag{t}[][][0.9]{$t$}
  \includegraphics[width=0.6\hsize]{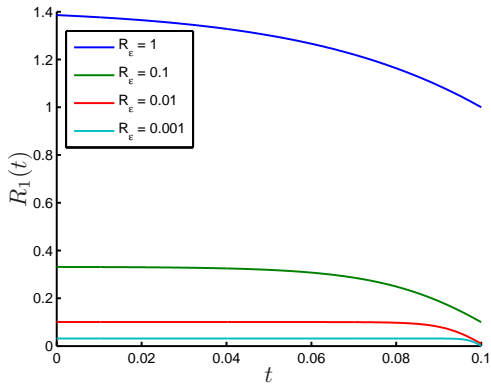}\\
  \caption{Optimal resistances $R_1(t)$ for several $R_\epsilon$.
  Here $R_2=1.0$, $C=0.1$, and $t_f=0.1$. For $t\ll t_f$ we have $R_1(t) \approx  \sqrt{R_2R_\epsilon + R_\epsilon^2}$.}\label{fig:R1t}
\end{figure}

Maximizing the amount of extracted work from the hot $R_2 C$ circuit within a finite time $t_f$ (which is minimizing the work  given to the same circuit) is an optimal control problem with the criterium
\begin{equation}
W^\star := - \min_u \E \int_0^{t_f} y(t)u(t)\, dt \quad \text{subject to } \eqref{eq:cap_heat_source},
\label{eq:opt_work}
\end{equation}
which we solve in the Appendix for all values of $t_f, R_{\epsilon}, R_2$ and $C$. Strikingly, the optimal controller is of the form $u=-y/R_1(t)$. In other words, the optimal heat engine is a zero-temperature time-varying resistance $R_1(t)$. A circuit representation of
the optimal heat engine is given in Figure~\ref{fig:LQR_circuit}-(b). For large time horizon, $R_1(t)$ assumes a nearly constant value $\sqrt{R_2 R_\epsilon + R_\epsilon^2}$ until roughly $t_f-\tau_2/(2\sqrt{1+R_2/R_\epsilon})$ where it converges exponentially fast to $R_\epsilon$, as illustrated in Figure~\ref{fig:R1t}. In case of an infinite horizon $t_f\rightarrow\infty$, $R_1$ takes the constant value $\sqrt{R_2 R_\epsilon + R_\epsilon^2}$.
The total work extracted by the optimal linear heat engine is
\begin{equation}
W^\star  =  \left(\frac{1}{2} + \frac{t_f}{\tau_2}\right)T -  \sqrt{\kappa}\left(1 +\frac{2t_f}{\tau_2}\right)T + O(\kappa),  \quad  \kappa=\frac{R_\epsilon}{R_2} \rightarrow 0, \,t_f\rightarrow \infty.
\label{eq:lqr_work}
\end{equation}

The power for large times is therefore $(1-2\sqrt{R_\epsilon/R_2})T/\tau_2$. Hence, a hot but small resistor $R_2$ has the potential to be a good source of work in the circuit Figure~\ref{fig:LQR_circuit}. Remark that the answer would be very different for a constant current source instead of random, where the maximum power transfer to $R_1$ is reached by an impedance matching $R_1\approx R_2$.

It has been assumed all the power leaving the circuit, $-y(t)u(t)$, is equal to the work extraction rate of the engine. One may wonder if there
is a physical device that can generate the optimal current $u=-y/R_1(t)$ while converting this power into useful mechanical work (for example) without any losses. This requires a device that emulates a time-varying resistor $R_1(t)$ of zero temperature. As the capacitor $C$ and $R_1(t)+R_\epsilon$ in Figure~\ref{fig:LQR_circuit}-(b) is simply an RC-circuit, it can be implemented with a lossless time-varying circuit, see Section~\ref{sec:time_varying_cap} and Figure~\ref{fig:LQR_circuit}-(c). The details of this and the corresponding energy
balance is the topic of the next subsection.

\subsection{Energy balance of the linear heat engine}
\label{sec:energyengine}

We focus here on the analysis of the constant $R_1$ case, which occurs for long or infinite time horizon $t_f$.
The model of the circuit in Figure~\ref{fig:LQR_circuit}-(b) then becomes
\begin{equation}
\frac{d}{dt} x = -\left(\frac{1}{\tau_1 + \tau_\epsilon}+\frac{1}{\tau_2}\right) x + \sqrt {\frac{2T}{\tau_2}}w,
\label{eq:lqr_implementation1}
\end{equation}
where $\tau_i=R_iC$. Naming $\tau^{-1}=(\tau_1 + \tau_\epsilon)^{-1}+\tau_2^{-1}$ the global time constant of the system, we can rewrite the model as
\begin{equation}
\frac{d}{dt} x = -\frac{1}{\tau} x + \sqrt {\frac{2T'}{\tau}}w,
\label{eq:lqr_implementation2}
\end{equation}
where $T'=T \tau / \tau_2$ acts as an `effective' temperature for which the model is identical to a simple Langevin's equation for an RC circuit in contact with one heat bath $T'$ and no work extraction. This reformulation allows a simple analysis of the energy and work balance of the system, as we now show. 

The balance of internal energy $U=\frac{1}{2}\E_x x^2$ is written
\begin{align*}
\dot U & = -\frac{2}{\tau}U + \frac{T'}{\tau}, \\
U(t) & = \frac{T'}{2} + \left(U(0)-\frac{T'}{2}\right)e^{-2t/\tau}.
\end{align*}
The expected power dissipated into $R_1+R_\epsilon$ is $-\E_x v_c(t)i(t)= \E_x \frac{x(t)}{\sqrt{C}}\frac{x(t)}{\sqrt{C}(R_1+R_\epsilon)}= 2 U(t)/(\tau_1+\tau_\epsilon)$, the fraction $\alpha:=\tau_1/(\tau_1+\tau_\epsilon)$ of which is useful work (i.e., dissipated into $R_1$):
\begin{align}
w(t) & = 2 U(t) \frac{\alpha^2}{\tau_1}=\alpha^2 \left(\frac{T'}{\tau}-\dot{U}(t)\right)\frac{\tau}{\tau_1} , \notag \\
W & = \int_0^{t_f}w\,dt = \alpha^2 \left(\frac{t_f}{\tau_1}T' - (U(t_f) - U(0)) \frac{\tau}{\tau_1}\right). \label{eq:RC_work_ext}
\end{align}
Assume the capacitor is initially in thermal equilibrium with $R_2$, i.e., $U(0)=T/2$, and is then also connected to $R_1+R_\epsilon$ of temperature zero. The capacitor will then exponentially fast reach a new thermal equilibrium with internal energy $U(t_f)\approx T'/2$ (for large enough $t_f$). The total work extracted during this relaxation becomes
\begin{align*}
W & = \alpha^2 T \left(\frac{t_f}{\tau_1+ \tau_\epsilon+\tau_2}+\frac{1}{2(1+\frac{\tau_1+\tau_\epsilon}{\tau_2})^2}\right).
\end{align*}
This expression can readily be maximized for $\tau_1$, which confirms that $\tau_1=\sqrt{\tau_\epsilon \tau_2 + \tau_\epsilon^2}$ (i.e., $R_1 = \sqrt{R_2R_\epsilon+R_\epsilon^2}$) is optimal and for small $\kappa=\tau_\epsilon/\tau_2$ recovers $W^\star$ as in (\ref{eq:lqr_work}).

The linear heat engine can be understood as a physical implementation of Maxwell's demon, in that it acts on a system in feedback control with the intent to extract work from the random fluctuations of a single heat source. Unlike the original Maxwell's demon \cite{maxwell1897theory, leff2010maxwell} portrayed as an intelligent being of some sort, our demon has an explicit implementation as a physical system, which is also the case in \cite{mandal2012work,strasberg2013thermodynamics}. While most demons explored in the literature act in discrete time with a finite set of actions (e.g. open or close a trap door), our demon acts in continuous time with a continuous set of actions. Although it can extract any desired amount of work from a single heat source, the demon does not formally contradict the Second Law because it does so in a non-cyclic way, as it includes in particular a time-varying capacitor with exponentially increasing capacitance (see Equation~\eqref{eq:RC-circuit} and around).

This analysis also reveals a perhaps troubling property of this linear heat engine.
The temperature of the capacitor in steady state converges to $T' = T \tau/\tau_2$. Since $\tau$ should be small to extract a large amount of work according to the above analysis, it indicates the optimal linear heat engine creates a large temperature gradient. This may seem to contradict an important message of thermodynamics: The most efficient heat engine (the Carnot heat engine) operates in quasi steady state avoiding finite temperature gradients and unnecessary entropy generation. This issue is further discussed in the next subsection.

\subsection{Finite-Time Carnot Heat Engine}
\begin{figure}[tb]
\centering
  \includegraphics[width=0.9\hsize]{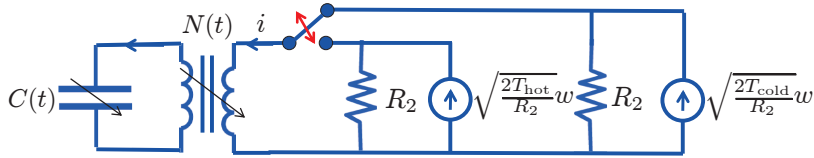}\\
  \caption{The heat engine circuit.}\label{fig:heat_engine_circuit}
\end{figure}

So far we have addressed the problem of extracting work from a hot heat bath of temperature $T$
using a zero-temperature resistor $R_1(t)$. This device can be implemented by a 
time-varying capacitor with exponentially increasing capacitance, see Section~\ref{sec:time_varying_cap} and in particular 
Equation~\eqref{eq:RC-circuit}.
From a practical perspective, it is clear
we cannot let this increase go on forever, and next we find a method to reset the capacitor to the initial state while
extracting net work.
Hence, we are interested in constructing a cyclic operation of the capacitor.
The classical way to operate heat engines is to introduce two heat baths of temperatures $T_{\textrm{cold}}<T_{\textrm{hot}}$.
This idea we will pursue next, based on the circuit in Figure~\ref{fig:heat_engine_circuit}, which is based upon Figure~\ref{fig:LQR_circuit}-(c). One should not confuse the time-varying capacitance $C(t)$ with the emulated constant capacitance $C$, part of the $R_1C$ emulated circuit. 

\begin{figure}[tb]
\centering
  \includegraphics[width=0.7\hsize]{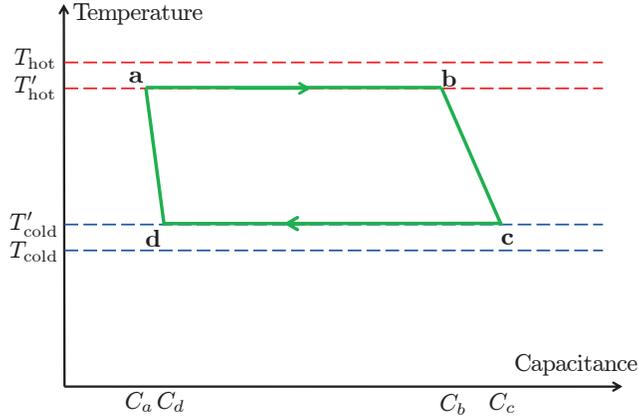}\\
  \caption{The finite-time Carnot cycle.  The capacitance is $C(t)$ as represented in Figure~\ref{fig:heat_engine_circuit}.}\label{fig:finite_time_Carnot}
\end{figure}

To reset the engine, we will construct a cycle resembling the Carnot cycle, see for example~\cite{Brockett+78}, and compute its efficiency. The cycle and its four legs are shown in Figure~\ref{fig:finite_time_Carnot}.
The time-varying capacitor first goes through an isothermal phase (\textbf{a}$\rightarrow$\textbf{b}) emulating a constant positive resistor $R_1^\textrm{hot}$  and a constant capacitance (temperature $T_\textrm{hot}'<T_\textrm{hot}$, time constant $\tau_1^\textrm{hot}$) of duration $t_\textrm{hot}$. The third leg is also an isothermal phase (\textbf{c}$\rightarrow$\textbf{d}) but implementing a negative resistor $-R_1^\textrm{cold}$ with a constant capacitance (temperature $T_\textrm{cold}'>T_\textrm{cold}$, time constant $\tau_1^\textrm{cold}>0$)
of duration $t_\textrm{cold}$. In these legs, the time-varying capacitance (the nonlinear input $v_1$) satisfies
\begin{equation*}
C_b = C_a e^{2t_\textrm{hot}/\tau_{1}^\textrm{hot}}, \quad C_d = C_c e^{-2t_\textrm{cold}/\tau_{1}^\textrm{cold}}.
\end{equation*}
The temperatures depend on the time constants as
\begin{equation*}
T_\textrm{hot}' = \frac{T_\textrm{hot}}{1+\tau_2/\tau_1^\textrm{hot}}, \quad T_\textrm{cold}' = \frac{T_\textrm{cold}}{1-\tau_2/\tau_1^\textrm{cold}},
\end{equation*}
which follow from models identical to \eqref{eq:lqr_implementation1} using $\tau_1 = \tau_1^\textrm{hot}$ and 
$\tau_1 = -\tau_1^\textrm{cold}$.
To close the cycle, two adiabatic legs (\textbf{b}$\rightarrow$\textbf{c}, \textbf{d}$\rightarrow$\textbf{a})
 are also introduced. These can be understood as stepwise instantaneous changes of the time-varying capacitance where the charge in the capacitor remains constant and we have
\begin{equation*}
T_\textrm{hot}'C_b = T_\textrm{cold}' C_c, \quad T_\textrm{cold}' C_d = T_\textrm{hot}' C_a,
\end{equation*}
in order to have a closed cycle.
From the above expressions, it is clear we must satisfy the constraints
\begin{equation}
t_\textrm{hot}/\tau_1^\textrm{hot} = t_\textrm{cold}/\tau_1^\textrm{cold}, \quad \tau_1^\textrm{cold}>\tau_2
\label{eq:closed_cycle_cond}
\end{equation}
to form stable closed cycles which can be repeated indefinitely.
In particular, in the cold phase when the resistance is negative, we must have $\tau_1^\textrm{cold}>\tau_2$ to have a finite temperature. If we decrease the capacitance at a too high rate, the temperature goes unbounded.

We can now compute the work and heat flows in the isothermal legs as (assuming $R_\epsilon=0$ and noting that $U$ is constant)
\begin{equation}
W_\textrm{hot} =  Q_\text{hot} =  T_\textrm{hot}'\frac{t_\textrm{hot}}{\tau_{1}^\textrm{hot}}, \quad W_\textrm{cold} =  Q_\textrm{cold} =  -T_\textrm{cold}'\frac{t_\textrm{cold}}{\tau_{1}^\textrm{cold}},
\end{equation}
using the work extraction formula \eqref{eq:RC_work_ext}. 
The work in the adiabatic legs are $\pm W_\textrm{ad}=\pm \frac{1}{2}(T_\textrm{hot}'-T_\textrm{cold}')$ and the heat flow is zero. Hence, work is extracted in the hot phase, and work is put back in to reset the capacitor in the cold phase. The efficiency of heat engines is typically defined as the net work over the cycle divided by the heat input. Here it becomes
\begin{equation}
\begin{aligned}
\eta  & = \frac{W_\textrm{hot}+W_\textrm{ad}+W_\textrm{cold}-W_\textrm{ad}}{Q_\textrm{hot}} = \frac{Q_\textrm{hot}+Q_\textrm{cold}}{Q_\textrm{hot}} \\ & = 1 - \frac{T_\textrm{cold}'}{T_\textrm{hot}'}  = 1 - \frac{T_\textrm{cold}}{T_\textrm{hot}} \frac{1+\tau_2/\tau_1^\textrm{hot}}{1-\tau_{2}/\tau_{1}^\textrm{cold}},
\label{eq:efficieny}
\end{aligned}
\end{equation}
where we have used the cycle condition \eqref{eq:closed_cycle_cond}. It is interesting that $\eta$ has the same form as the Carnot heat engine efficiency, except that one should use the effective temperatures $T_\textrm{hot}'$ and $T_\textrm{cold}'$ instead of $T_\textrm{hot}$ and $T_\textrm{cold}$. In particular, net work is only obtained if $T_\textrm{cold}'<T_\textrm{hot}'$ which puts constraints on how to operate the engine.

The larger the difference between $T_\textrm{hot}'$ and $T_\textrm{cold}'$, the higher the efficiency. This is obtained by making the ratios $\tau_2/\tau_1^\textrm{hot}$ and $\tau_{2}/\tau_{1}^\textrm{cold}$ small. In fact, we can come arbitrarily close to the Carnot heat engine efficiency by making $R_1^\textrm{hot}$ and $R_1^\textrm{cold}$ large relative to $R_2$. It is also interesting to note that the efficiency does not depend on the period time $t_\textrm{hot}+t_\textrm{cold}$ and the emulated capacitance $C$.

Another quantity of interest is the mechanical power defined and given by
\begin{equation}
\begin{aligned}
\bar{w} & := \frac{W_\textrm{hot}+W_\textrm{ad}+W_\textrm{cold}-W_\textrm{ad}}{t_\textrm{hot}+t_\textrm{cold}} \\ & = \frac{1}{\tau_1^\textrm{hot}+\tau_1^\textrm{cold}}\left(\frac{T_\textrm{hot}}{1+\tau_2/\tau_1^\textrm{hot}} - \frac{T_\textrm{cold}}{1-\tau_2/\tau_1^\textrm{cold}}\right).
\end{aligned}
\label{eq:ave_power}
\end{equation}
The power is the net work averaged over a cycle, and can be made
arbitrarily large by choosing the emulated capacitance $C$ small. The efficiency $\eta$ can be made large by choosing $R_1^\textrm{hot}$ and $R_1^\textrm{cold}$ large. Therefore power and efficiency can be simultaneously high. Note, however,
that if there is a lower bound on the time constant $\tau_2=R_2C$, for instance, then there is a trade-off between efficiency  and power. It is a simple calculation to find that the optimal power is reached for
\begin{equation}
\frac{\tau_1^{\textrm{cold}}}{\tau_2}=\frac{\tau_1^{\textrm{hot}}}{\tau_2}=\frac{R_1^{\textrm{cold}}}{R_2}=\frac{R_1^{\textrm{hot}}}{R_2}= \frac{\sqrt{T_{\textrm{hot}}}+\sqrt{T_\textrm{cold}}}{\sqrt{T_{\textrm{hot}}}-\sqrt{T_\textrm{cold}}}
\label{eq:opt-tau}
\end{equation}
A topic of finite-time thermodynamics is to characterise the maximum power cycle, often in terms of the thermal conductivity $k$ between the bath and the system. Here we may identify the thermal conductivity to $q/(T-T')=1/\tau_2$, for both baths. With this identification, we recover the maximum power $(\sqrt{T_\textrm{hot}} - \sqrt{T_\textrm{cold}})^2/4\tau_2$, as predicted by the classical Orlov-Berry formula \cite{orlov1990power} and the corresponding Chambadal-Novikov efficiency $1-\sqrt{T_\textrm{cold}/T_\textrm{hot}}$ \cite{chambadal1957centrales, novikov1958efficiency}.

Let us conclude by discussing the relation to the optimal linear heat engine and the issue raised in the end of Section~\ref{sec:energyengine}.
For this engine it was optimal to choose $R_1^\textrm{hot}=\sqrt{R_\epsilon^2 + R_2R_\epsilon}\rightarrow 0$ as $R_\epsilon \rightarrow 0$. This indeed gives the maximum possible power $W_\textrm{hot}/t_\textrm{hot}$ during the hot phase since $T_\textrm{hot}'\rightarrow 0$. But if we take the resetting of the capacitor using a cold heat source into account, the net efficiency is very bad, and even negative since $\tau_1^\textrm{cold}>\tau_2$. A negative efficiency means that it requires more work to reset the engine than was extracted in the hot phase. Finally,
note that the optimal linear engine assumed a fixed capacitance $C$ and was only optimized for the hot phase.

\section{Conclusion}

We have shown in this paper how even the most classical linear control theory can help us explore the fine performance of finite-time heat engines. We believe that this is only an example on how a better integration of existing control-theoretic tools, e.g.,  Kalman filtering, port-Hamiltonian theory, passivity theory, information-theoretic techniques in control,  etc. may be better integrated with statistical physics, in order to explore the fundamental limits to work extraction, actuation, measurement, or computation---even in a nonlinear context, which is conveniently formalised in the port-Hamiltonian framework.

\section*{Acknowledgments}
Stimulating conversations with John C. Doyle, Sanjoy Mitter, Roger Brockett and Christopher Jarzynski are gratefully acknowledged.
The work of Henrik Sandberg is supported by the Swedish Research Council under
Grants 2007-6350 and 2009-4565.
Jean-Charles Delvenne is supported by the Concerted Research Action (ARC) ``Large Graphs and Networks'' of the French
Community of Belgium, and by the Belgian Programme on Interuniversity Attraction Poles (IAP DYSCO) initiated by the Belgian Federal
Science Policy Office. He is also a member of Namur Complex Systems Center (naXys, Belgium).

\section*{References}

\appendix
\section{Proofs}

\subsection{Proof of Proposition \ref{prop:approx}}

That $b(t)c(t)> 0$ is a necessary condition is seen by applying the impulse input $u(t)=\delta(t-t_0)$, for arbitrary $t_0\geq 0$, to both \eqref{eq:captransdev} and \eqref{eq:genlin}. We obtain the outputs
$y(t_0^+)=v_2(t_0)^2/v_1(t_0)$ and $y(t_0^+)=b(t_0)c(t_0)$, respectively. These can only be made equal if $b(t_0)c(t_0)> 0$ given that $v_1(t_0)>0$ and $v_2(t_0)> 0$.

To show sufficiency, we construct explicit nonlinear inputs as follows. First apply the coordinate transformation $p = \sqrt{\frac{b}{c}} x$ to \eqref{eq:genlin} (which leaves the input-output relation invariant), and we obtain
\begin{equation*}
\begin{aligned}
\dot{x} & = \left(a - \frac{\dot b}{2b} + \frac{\dot c}{2c}\right) x + \sqrt{bc} u, \quad
y  = \sqrt{bc} x.
\end{aligned}
\end{equation*}
Hence, the time-varying capacitance $v_1(t)$ in \eqref{eq:captransdev} should satisfy the differential equation
\begin{equation*}
    -\frac{\dot v_1}{2v_1} = a - \frac{\dot b}{2b} + \frac{\dot c}{2c},
\end{equation*}
with the solution
$v_1(t) = \exp \left[-2\int_0^t a(s) ds \right] \frac{b(t)c(0)}{c(t)b(0)}v_1(0)$, where $v_1(0)>0$ is arbitrary.
Finally, the turns ratio of the transformer should be chosen as $v_2(t) = \sqrt{b(t)c(t)v_1(t)}$. This concludes the proof.

\subsection{The optimal finite-time extraction of work from a resistance}
\begin{figure}[tb]
\centering
    \psfrag{S(t)}[][][0.9]{$S(t)$}
    \psfrag{t}[][][0.9]{$t$}
  \includegraphics[width=0.6\hsize]{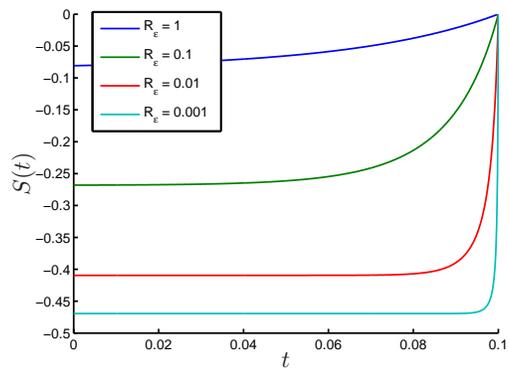}\\
  \caption{Solutions to the Riccati equation for several $R_\epsilon$. Here $R_2=1.0$, $C=0.1$, and $t_f=0.1$.}\label{fig:St}
\end{figure}

The optimization problem (\ref{eq:opt_work}) is a particular case of so called Linear Quadratic Regulation theory, which seeks to find a controller for a linear system that optimizes an integral over time of a quadratic function of the states and inputs of the resulting trajectories of the system, see, for example \cite{Astrom}. Its solution is characterized by the control Riccati equation
\begin{equation*}
\dot S = \frac{2}{\tau_2}S + \frac{1}{R_\epsilon C} \left(S+\frac{1}{2}\right)^2, \quad S(t_f) = 0,
\end{equation*}
with solution
\begin{align*}
    S(t) & = S_0 \frac{e^{\lambda(t_f-t)}-1}{e^{\lambda(t_f-t)}-\Lambda^2}, \quad S_0 = -\frac{1}{2}(\sqrt{\kappa}-\sqrt{\kappa + 1})^2,  \\
    \kappa & = \frac{R_\epsilon}{R_2},\quad \lambda = \frac{2}{\tau_2}\sqrt{1+1/\kappa}, \quad \Lambda = \frac{\sqrt{\kappa}-\sqrt{\kappa + 1}}{\sqrt{\kappa} + \sqrt{\kappa +1}}.
\end{align*}
When the time interval $[0,t_f]$ becomes large ($t_f > \frac{\tau_2}{2\sqrt{1+1/\kappa}}$), we have that
\begin{equation*}
    S(0) \rightarrow S_0 = -\frac{1}{2}(\sqrt{\kappa}-\sqrt{\kappa + 1})^2 = -\frac{1}{2} + \sqrt{\kappa}\ + O(\kappa), \quad \kappa \rightarrow 0.
\end{equation*}
Some example trajectories of $S$ are shown in Figure~\ref{fig:St}. As can be seen, they converge to the steady-state value $S_0$ exponentially fast when $t$ decreases from $t_f$. The optimal input current to inject into the circuit to extract the maximum amount of work is given by the feedback
\begin{equation}
u =-\frac{S(t)+1/2}{R_\epsilon \sqrt{C}} x =-\frac{y}{R_1(t)},
\label{eq:opt_feedback}
\end{equation}
where
$$R_1(t)=R_\epsilon \frac{1/2 - S(t)}{1/2 + S(t)}.$$

If $t_f$ is large, then we can approximate $S(t)$ by $S_0$ for most of the time, and find
$$
R_1=\sqrt{R_\epsilon R_2 + R_\epsilon^2}.
$$
Observe that for small $R_\epsilon$, the optimal resistance is the geometric mean of the
heat supply resistance and the loss resistance.

The maximum amount of work \eqref{eq:opt_work} that can be extracted is given by (see \cite{Astrom})
\begin{equation*}
W^\star  = -S(0)T - \int_0^{t_f} \frac{2T}{\tau_2}S(t)\,dt
\rightarrow -T\left( 1 - \frac{2t_f}{\tau_2}  \right) S_0 ,  \quad t_f\rightarrow \infty
\end{equation*}
which in the limit of small $\kappa$ gives \eqref{eq:lqr_work}. The optimal amount of extracted work for large $t_f$ roughly scales as $T/\tau_2$. Hence, a warm but small resistor $R_2$ has the potential to be a good source of work in the circuit in Figure~\ref{fig:LQR_circuit}.

\end{document}